\documentclass[aps,twocolumn,floats,prl,nofootinbib,superscriptaddress]{revtex4-1}

\usepackage[dvips]{graphicx} %
\usepackage{graphicx,amsmath,amsfonts,amssymb,slashed}
\usepackage{bbold,wasysym}

\usepackage{multirow, booktabs, dcolumn}

\usepackage{tabularx}
\usepackage{aas_macros}
\usepackage{simplewick}
\usepackage{hyperref}  

\usepackage[usenames,dvipsnames]{xcolor} 

\usepackage{soul}

\definecolor{RedWine}{rgb}{0.743,0,0}
\definecolor{RoyalBlue}{rgb}{0.25,.41,.88}

\setstcolor{Blue}

\renewcommand{\sec}{\ensuremath{\mathrm{s}}}

\newcommand{\keV}{\ensuremath{\mathrm{keV}}}
\newcommand{\MeV}{\ensuremath{\mathrm{MeV}}}
\newcommand{\GeV}{\ensuremath{\mathrm{GeV}}}
\newcommand{\TeV}{\ensuremath{\mathrm{TeV}}}

\newcommand{\deut}{\ensuremath{\mathrm{D}}}

\newcommand{\hyd}{\ensuremath{\mathrm{H}}}

\newcommand{\hef}{\ensuremath{{}^4\mathrm{He}}}
\newcommand{\het}{\ensuremath{{}^3\mathrm{He}}}
\newcommand{\lisx}{\ensuremath{{}^6\mathrm{Li}}}
\newcommand{\lisv}{\ensuremath{{}^7\mathrm{Li}}}

\newcommand{\bes}{\ensuremath{{}^7\mathrm{Be}}}

 \def\lsim{\mathrel{\rlap{\lower4pt\hbox{\hskip1pt$\sim$}}
    \raise1pt\hbox{$<$}}}
\def\gsim{\mathrel{\rlap{\lower4pt\hbox{\hskip1pt$\sim$}}
    \raise1pt\hbox{$>$}}}

\usepackage{tikz}

\usetikzlibrary{positioning,arrows,patterns}
\usetikzlibrary{decorations.markings}
\usetikzlibrary{calc}

\usetikzlibrary{positioning,arrows,patterns}
\usetikzlibrary{decorations.markings}
\usetikzlibrary{calc}

\tikzset{
    photon/.style={decorate, decoration={snake}, draw=black},
    vector/.style={decorate, decoration={snake}, draw},
	provector/.style={decorate, decoration={snake,amplitude=2.5pt}, draw},
	antivector/.style={decorate, decoration={snake,amplitude=-2.5pt}, draw},
    fermion/.style={draw=black, postaction={decorate},
        decoration={markings,mark=at position .55 with {\arrow[draw=black]{>}}}},
    fermionbar/.style={draw=black, postaction={decorate},
        decoration={markings,mark=at position .55 with {\arrow[draw=black]{<}}}},
    fermionnoarrow/.style={draw=black},
    gluon/.style={decorate, draw=black,
        decoration={coil,amplitude=4pt, segment length=5pt}},
    scalar/.style={dashed,draw=black, postaction={decorate},
        decoration={markings,mark=at position .55 with {\arrow[draw=black]{>}}}},
    scalarbar/.style={dashed,draw=black, postaction={decorate},
        decoration={markings,mark=at position .55 with {\arrow[draw=black]{<}}}},
    scalartwo/.style={dotted,draw=black, postaction={decorate},
        decoration={markings,mark=at position .55 with {\arrow[draw=black]{>}}}},
    scalartwobar/.style={dotted,draw=black, postaction={decorate},
        decoration={markings,mark=at position .55 with {\arrow[draw=black]{<}}}},
    scalarnoarrow/.style={dashed,draw=black},
    electron/.style={draw=black, postaction={decorate},
        decoration={markings,mark=at position .55 with {\arrow[draw=black]{>}}}},
	bigvector/.style={decorate, decoration={snake,amplitude=4pt}, draw},
    vertex/.style={draw,shape=circle,fill=black,minimum size=1pt,inner sep=0pt},
}

\begin{document}

\title{A light particle solution to the cosmic lithium problem}

\author{Andreas Goudelis}
\affiliation{Institute of High Energy Physics, Austrian
     Academy of Sciences, Nikolsdorfergasse 18, 1050 Vienna,
     Austria}
\author{Maxim Pospelov}
\affiliation{Perimeter Institute, Waterloo, Ontario N2L 2Y5, Canada}
\affiliation{Department of Physics and Astronomy, University of Victoria, Victoria, BC, V8P 5C2, Canada}
\author{Josef Pradler}
\affiliation{Institute of High Energy Physics, Austrian
     Academy of Sciences, Nikolsdorfergasse 18, 1050 Vienna,
     Austria}

\begin{abstract}
We point out that the cosmological abundance of ${}^7$Li can be
reduced down to observed values if during its formation Big Bang
Nucleosynthesis is modified by the presence of light electrically
neutral particles~$X$ that have substantial interactions with
nucleons. We find that the lithium problem can be solved without
affecting the precisely measured abundances of deuterium and helium if
the following conditions are satisfied: the mass (energy) and lifetimes of such
particles are bounded by $ 1.6~{\rm MeV}\leq m_X (E_X) \leq 20~{\rm MeV}$
and $ {\rm few}~100~{\rm s} \lesssim \tau_X \lesssim 10^4~{\rm s}$,
and the abundance times the absorption cross section by either
deuterium or ${}^7$Be are comparable to the Hubble rate, $n_X
\sigma_{\rm abs} v \sim H$, at the time of ${}^7$Be formation.  We
include $X$-initiated reactions into the primordial nucleosynthesis
framework, observe that it leads to a substantial reduction of the
freeze-out abundances of ${}^7$Li+${}^7$Be, and find specific model
realizations of this scenario.  Concentrating on the
axion-like-particle case, $X=a$, we show that all these conditions can
be satisifed if the coupling to $d$-quarks is in the range of $f_d^{-1}
\sim {\rm TeV}^{-1}$, which can be probed at intensity frontier
experiments.
\end{abstract}

\maketitle

\paragraph{Introduction.} Big Bang Nucleosynthesis (BBN) is a
cornerstone of modern cosmology~\cite{Fields:2014uja,Cyburt:2015mya}.
Its success rests on the agreement among the observationally inferred
and predicted primordial values for the deuterium and helium
abundances. In particular, the latest measurements of the deuterium
abundance,
$(\deut/\hyd)_{\rm obs} = (2.53 \pm 0.04) \times
10^{-5}$~\cite{Cooke:2013cba},
are in remarkable accord with BBN predictions under standard
cosmological assumptions, and using the baryon-to-photon
ratio---precisely measured via the anisotropies in the cosmic
microwave background (CMB)~\cite{Ade:2015xua}---as an input. However,
the BBN success is not complete: the predicted value of the lithium
abundance~\cite{Cyburt:2015mya},
$(\lisv/{\rm H})_{\rm BBN}=(4.68 \pm 0.67) \times 10^{-10}$, is
significantly higher, by a factor of $\sim (2-5)$, than the value
inferred from the atmospheres of PopII stars,
$(\lisv/{\rm H})_{\rm obs}= (1.6 \pm 0.3) \times 10^{-10}$~\cite{Sbordone:2010zi}. What
prevents this discrepancy, known as the \textit{cosmological lithium
  problem}, from becoming a full-blown crisis for cosmology is the
questionable interpretation of $(\lisv/{\rm H})_{\rm obs}$ as being
the truly primordial value, unaltered by subsequent astrophysical
evolution. Indeed, several astrophysical mechanisms of how the
reduction of lithium may have come about have been proposed (see, {\em
  e.g.} \cite{Korn:2006tv,2015MNRAS.452.3256F}), none of which resolve the
problem completely. Thus, {\em New Physics} (NP) scenarios, such as
modifications of standard BBN, can be entertained as solutions to this
long-standing discrepancy.

The (over)abundance of lithium is ultimately related to the excessive
production of the \bes\ isotope, that radiatively decays to \lisv\
during the post-BBN evolution. Its reduction occurs at $T\gsim 25\keV$
via the sequence of neutron capture in the $\bes(n,p)\lisv$ reaction,
followed by $\lisv(p,\alpha)\hef$.  For a while, NP scenarios
supplying {\em extra} neutrons, thereby reducing the \lisv+\bes\
abundance \cite{Reno:1987qw,Kawasaki:2004yh,Jedamzik:2006xz}, were
considered to be attractive solutions to the lithium problem. However,
in light of the latest (D/H) measurements~\cite{Cooke:2013cba}, {\em
  any} such solution is strongly disfavored
\cite{Coc:2014gia,Kusakabe:2014ola} as extra neutrons lead to the
overproduction of deuterium, quite generically resulting in
$(\deut/\hyd)_{\rm BBN} > 3 \times 10^{-5}$, far from the allowed
range.  This excludes a variety of models with late decays of
electroweak-scale particles, including many supersymmetric
scenarios. Nevertheless, isolated cases of NP models, typically
involving sub-GeV particles, can reduce lithium while keeping
deuterium and helium consistent with
observations~\cite{Pospelov:2010cw,Poulin:2015woa}. We also note that
 BBN catalyzed by the presence of negatively charged weak-scale
particles \cite{Pospelov:2007mp,Bird:2007ge,Kusakabe:2014moa} still
has potential for reducing the \bes\ abundance.

\begin{figure}[hb]
  \centering
\begin{tikzpicture}[line width=1.1 pt, scale=1.5]
\coordinate (v1) at (1.2,-1);
\coordinate (v2) at (1.2,-0.95);
\coordinate (v3) at (1.2,-1.05);
\coordinate[label=left :$X$]  (e1) at (0,-0.28);
\coordinate[label=left :{$\bes\, (\deut) $}]  (e2) at (0,-0.98);
\coordinate[label=right :{$\hef\, (p)$}] (e4) at (2.4,-0.98);
\coordinate[label=right :{$\het\,(n)$}] (e3) at (2.4,-0.3);
\draw[scalar] (0,-0.3) -- (v1) ;
\draw[fill] (v1) circle [radius=0.1];
\draw[fermionnoarrow] (0,-0.95) -- (v2);
\draw[fermionnoarrow] (0,-1.05) -- (v3);
\draw[fermion] (v3) -- (2.4,-1.05) ;
\draw[fermion] (v2) -- (2.4,-0.3) ;
\end{tikzpicture}
\caption{\small Spallation of a nucleus due to absorption of a bosonic
  state $X$. }
  \label{fig:absorp}
\end{figure}
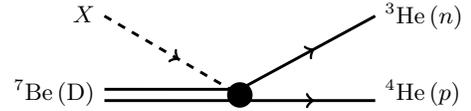

In this {\em Letter} we suggest a new mechanism for selectively
reducing the lithium abundance, while keeping other BBN predictions
intact.  \bes\ is formed in the narrow temperature range from 60 to
40~keV, after deuterium- and during \het-formation, in a rather slow,
sub-Hubble rate reaction $ \het(\alpha,\gamma)\bes$. This is why its
abundance is very small, $(\bes/\het)\ll 1$, and it contrasts with
other nuclear reactions responsible for \hef, \het, D, which remain
very fast in that temperature window.  Therefore, if BBN is modified
by a new light and meta-stable neutral particle $X$ that has direct
interactions with nucleons and can react as in Fig.~\ref{fig:absorp},
{\em either} with \bes\ or deuterium (or both)  via
\begin{equation}
{\rm R1}:~ \bes(X,\alpha)\het;~~{\rm R2}:~ {\rm D}(X,p)n
\label{reactions} 
\end{equation} 
at $T\sim 50~\keV$, then one should expect that the \bes\ (and
consequently the observed \lisv) abundance will be reduced. Most
importantly, if reactions R1 and R2 occur relatively early,
$T>10~\keV$, and the energy carried by the $X$ particle is below the
\hef\ binding energy, the helium and deuterium abundance will not be
altered in a significant way, as neutrons generated in R2 will be
incorporated back to deuterium via the process $p(n,\gamma){\rm D}$
that remains faster than neutron decay down to temperatures of
$T\sim 10 ~\keV$. Note that $X$ cannot be a light Standard Model
particle; non-thermal photons at these temperatures are quickly
degraded in energy below nuclear binding thresholds, and neutrinos
have too small an interaction rate.

In the remainder of this paper, we show that these qualitative
expectations are supported by detailed BBN calculations.  We determine
the required properties of $X$, provide concrete particle physics
realizations, and point out experimental avenues to test the proposed
scenarios.

\paragraph{New light metastable particles during BBN.} Light, very
weakly coupled particles $X$ can  selectively affect BBN processes if
their number density is large, but their energy density remains
subdominant to that of photons. Therefore, as a guideline, we shall
assume that their number density during BBN satisfies the bound
\begin{equation}
n_b \lsim n_X < \frac{T}{E_X}\times n_\gamma,
\label{abundance}
\end{equation}
where $E_X$ is the energy carried by these particles (and $E_X=m_X$
for the non-relativistic case).  Since the respective baryon and
photon number densities $n_b$ and $n_{\gamma}$ are widely different,
$n_b/n_\gamma = 6.1\times 10^{-10}$~\cite{Ade:2015xua}, the abundance
of $n_X$ (\ref{abundance}) can vary in a rather large range. We
distinguish two different scenarios.  {\em Scenario A} assumes that
$X$ is non-relativistic, with mass in the range from $1.6$ to 20 MeV,
and it participates in the reactions (\ref{reactions}) before decaying
either to Standard Model (SM) radiation, or to a beyond-SM radiation
species. {\em Scenario B} assumes that there is an inert, almost
non-interacting neutral progenitor particle $X_p$ that decays to
(nearly) massless states $X$ which participate in the nuclear
reactions before being red-shifted below nuclear reaction
thresholds. For the two-body decay, $X_p\to XX$, the mass $X_p$ must
lie in the range from $3.2$ to 40~MeV, and the mass of $X$ should be
less than $\sim 1$~eV (to avoid hot dark matter constraints.) The
upper mass bound in both scenarios ensures that \hef\ is not directly
affected by $X$-induced splitting.

We modify our BBN code~\cite{Pospelov:2010hj} to include the effects
of $X$ particles. In the following we expose the relevant physics by
using Scenario~A for which we add the parameters
$\{ m_X, \tau_X, n_X/n_b, \sigma_{\rm Be} v, \sigma_{\rm D} v \}$ to
the code, where $n_X$ stands for the initial (un-decayed) abundance of
$X$ and $ \sigma_{\rm Be} v, \sigma_{\rm D} v$ are the respective
reaction cross sections for (\ref{reactions}). We assume that they are
dominated by the $s$-wave of initial particles, for which they become
temperature-independent parameters. The reactions with $A=3$ elements,
{\em e.g.} $\het(X,p){\rm D}$, are generically less important and, in
the interest of concision, we avoid them altogether by taking
$2.2~{\rm MeV} < m_X<5.5$~\MeV. 
We note in passing, though, that $m_X > 5\,\MeV$ may be beneficial since
$ \bes (X,p)\lisx$ opens as an additional depleting channel.
Note that the assumed small couplings
of $X$ and large abundances (\ref{abundance}) make the reverse
reactions, {\em e.g.} $n(p,X){\rm D}$, negligible.
\begin{figure}
\centering
\includegraphics[width=\columnwidth]{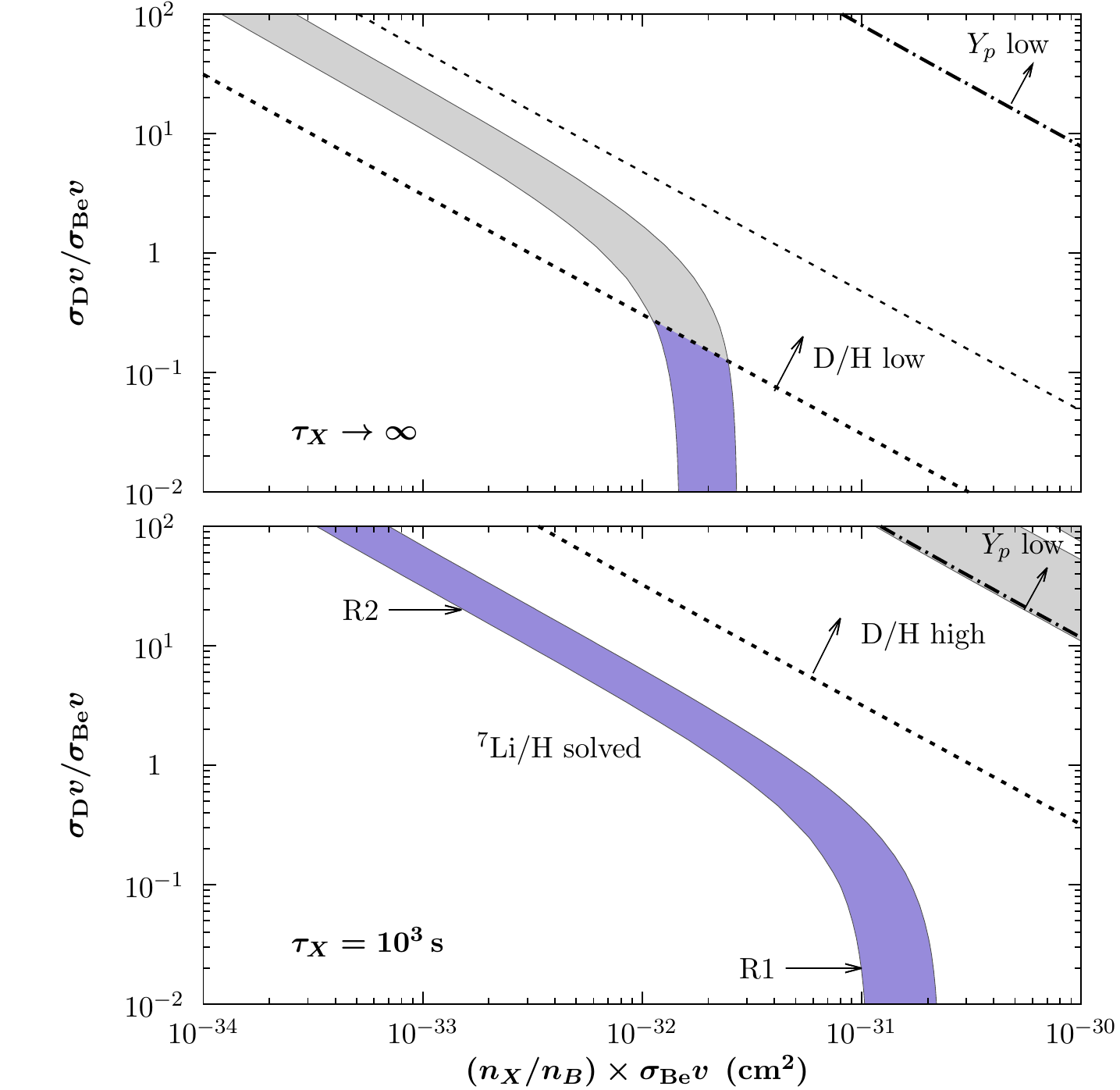}
\caption{\small The contours of light element abundances as a function
  of the two reaction rates R1 and R2 in Scenario A, for
  $\tau_X \gg t_{\rm BBN}$ (top panel), and $\tau_X =10^3{\rm s}$
  (lower panel); $\sigma_D$ is constant along the dotted lines. Inside
  the shaded regions, the lithium problem is solved.}
\label{fig:BBN}
\end{figure}

\begin{figure}
\centering
\includegraphics[width=\columnwidth]{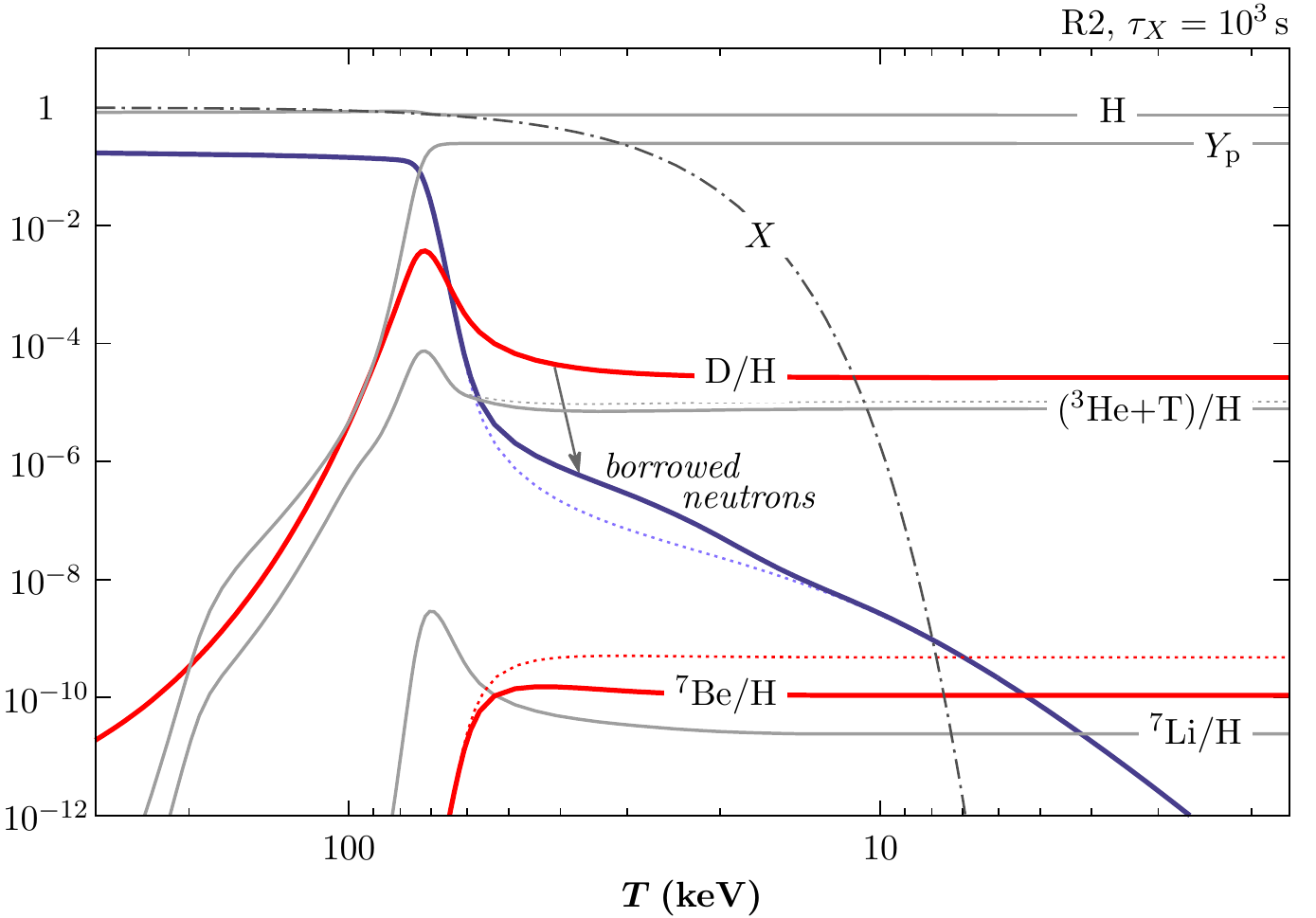}
\caption{\small Temperature evolution of elemental abundances, with
  BBN modified by R2, initiated by $X$ with $\tau_X=10^3$~s and
  $\frac{n_X}{n_b} \sigma_{\rm D} v = 5\times 10^{-32}~{\rm
    cm^2}$.
  The temporary increase in $n$ leads to the suppression of \bes\, but
  does not affect $(\deut/\hyd)_{\rm BBN}$. The dotted lines correspond 
  to the prediction of standard BBN.}
\label{fig:evol}
\end{figure}

The results of our calculations are presented in
Fig.~\ref{fig:BBN}. The dark shaded regions correspond to reaction
rates that reduce lithium to the range
$\lisv/\hyd = (1-2)\times 10^{-10}$ without affecting other
elements. In the top panel, the lifetime of $X$ is taken to be large
with respect to the cosmic time at BBN and, consequently, the late
reaction R2 reduces the deuterium abundance too much, unless
$\sigma_{\rm Be} > 10 \sigma_{\rm D}$.  Such a hierarchy of cross
sections would require additional tuning of the properties of $X$.  In
contrast, lifetimes around $10^3$ seconds (lower panel) allow for a {\em
  generic solution} to the lithium problem, without altering deuterium
beyond the observational bounds. In the vertical part of the shaded
band,  corresponding to small values of $\sigma_D$, \bes\ is directly depleted via R1, while in the diagonal part
 $\sigma_{\rm Be}$ is small and \bes\ reduction is achieved via neutrons generated through R2.  Note
that contrary to models of decaying weak-scale particles these are not
extra neutrons, but {\em borrowed} ones, that return to deuterium via
the fast reaction $p(n,\gamma){\rm D}$. Thus for $\tau_X\sim 10^3$~s,
the preferred R1 or R2 reaction rates solving the \lisv\
overproduction problem are
\begin{align}
\text{R1:}& \quad (n_X/n_b) \times \sigma_{\rm Be} v \simeq (1-2)\times 10^{-31}~{\rm cm^2}, \quad \text{or} \nonumber \\ 
\text{R2:}& \quad (n_X/n_b) \times \sigma_{\rm D} v \simeq (3-7)\times 10^{-31}~{\rm cm^2}. 
\label{sigma} 
\end{align}
The observational constraints in Fig.~\ref{fig:BBN} are
$2.45 \times 10^{-5} \leq \deut/\hyd \leq 3\times 10^{-5}$ (lower
limit nominal $2\sigma$ from \cite{Cooke:2013cba}; upper limit
conservative) and $Y_p \geq 0.24$; also shown is the unlabeled
D/H contour $10^{-5}$.
The effect of the ``borrowed'' neutrons resulting from R2 is shown in
Fig.~\ref{fig:evol}.

The absorption rates in~(\ref{sigma}), determined for $\tau_X$ on the
order of $10^3$ seconds, are comparable to the Hubble rate during
$^7$Be synthesis as should be expected from the NP-modified BBN
scenarios that achieve a factor of $O({\rm few})$ reduction of the
beryllium abundance. Short $X$ lifetimes, $\tau_X \ll 10^4$s, have the
additional benefit of reducing the sensitivity to visible decays of
$X$ to $\gamma\gamma$ or $e^+e^-$, as BBN is largely insensitive to
electromagnetic energy injections at early times (see, {\em e.g.}
\cite{Pospelov:2010hj}).  Similar solutions can be found in Scenario
B, where $\tau_{X_p}$ should be chosen in a similar range, while the
R1/R2 reaction rates will receive an additional temperature dependence
due to the redshift of $E_X$. A full scan of the viable parameter
space will be presented in a more detailed
publication~\cite{GPPinprep}.

\paragraph{Model realization: couplings, cross sections, abundance.}
The respective ranges (\ref{abundance}) and (\ref{sigma}) for the
abundances and reaction rates suggest a typical size for the R1 and/or
R2 cross sections.  If we choose $X$-particles of $ \sim 5~\MeV$ mass
(or energy) to contribute 1\% of the photon energy density at
$T=50~\keV$, we arrive at
$ \sigma_{\rm abs} v \sim 10^{-38}~{\rm cm}^2$. This is much smaller
than the typical ({$\sim$mbn) range for photonuclear reactions, and
  much larger than typical weak scale cross sections
  $\sim G_F^2 (E_X)^2$. Yet, the lifetimes of $X$ particles are
  commensurate with $\beta$-decay lifetimes, implying very small
  couplings to electrons, photons and neutrinos. It is then clear that
  only selected particle physics models can simultaneously account for
  (\ref{abundance}), (\ref{sigma}) and $\tau_X\sim10^3$s.

  A variety of models involving light, weakly interacting particles
  have been extensively studied in recent years~\cite{Essig:2013lka},
  including axions, axion-like particles (ALPs), and ``dark'' vectors.
  The MeV-mass range has been independently motivated as an ideal
  range for the force carrier that mediates dark matter
  self-interactions~\cite{Loeb:2010gj,Kaplinghat:2015aga}, as well as
  its interactions with the SM. Here we provide ``proof of existence''
  of models that satisfy the requirements on $\tau_X$,
  $\sigma_{\rm abs}$ and $n_X$ derived from our BBN analysis.

  If $X$ is massive (Scenario A), its decay to leptons will scale as
  $\Gamma_{e^+e^-}\propto m_X g_e^2/(4\pi)$. Given a lifetime of
  $10^3$s, the coupling to electrons would have to be smaller than
  $g_e \lsim 10^{-12} $.  At the same time, the coupling $g_N$ to
  nucleons will have to be much larger, pointing to ``leptophobic''
  models of light particles. Models with ``dark photons''
  \cite{Essig:2013lka} would hence not provide viable solutions, while
  models based on gauged baryon number $U(1)_B$
  \cite{Pospelov:2011ha,Tulin:2014tya} would have to be tuned to
  suppress the loop-induced couplings to leptons.  Models based on
  so-called axion-like particles represent a better candidate,
  and below we outline their main features. We consider a model where
  the $X$ particle is an ALP $a$ which interacts mainly with down-type
  quarks.  To avoid strong constraints from the flavour-violating $K$
  and $B$ meson decays, mediated by the top-$W$ loop, the coupling to
  up-type quarks is assumed to be suppressed.  We note in passing that
  such construction can be UV-completed by using multiple Higgs bosons
  and an interaction $H_uH_d\exp\{i a/f_a\}$, that gives $f_d\gg f_u$
  when $\langle H_u\rangle \gg \langle H_d\rangle $.  Going from the
  quark-ALP to the meson/nucleon-ALP interaction, we obtain the most
  important interactions with neutrons, protons and pions.
\begin{align}
{\cal L}_{aq} & = \frac{\partial_\mu a}{f_d} \bar d \gamma_\mu\gamma_5 d \quad \Longrightarrow  \nonumber 
\\  {\cal L}_{a\pi N} & =\frac{\partial_\mu a}{f_d}\left[ f_\pi \partial_\mu \pi^0  +
\frac43  \bar n\gamma_\mu\gamma_5 n - \frac13 \bar p\gamma_\mu\gamma_5 p\right].
\label{ALPd}
\end{align}
We have used a naive quark model estimate for the spin content of the
nucleons, and $f_\pi = 93$~\MeV.  The kinetic mixing of the two
scalars results in a small admixture of $\pi^0$ to an on-shell $a$,
with the mixing angle $\theta = (f_\pi/f_d)\times (m_a^2/m_\pi^2)$,
and induces the decay $a\to \gamma\gamma$. Upon appropriate rescaling,
$\Gamma^{a}_{\gamma\gamma} \simeq \theta^2 \left(
  \frac{m_a}{m_\pi}\right)^3 \Gamma^{\pi^0}_{\gamma\gamma} $,
which gives the lifetime in the right ballpark for $f_d \sim $\TeV\
and $m_a\sim$ 5~MeV.  The coupling of $a$ to the $\gamma_\mu\gamma_5$
nucleon current leads to the nonrelativistic Hamiltonian proportional
to nucleon helicities. To estimate the absorption cross sections we
follow the method of \cite{Pospelov:2008jk} that relates the ALP
absorption to the photoelectric effect in the dipole (E1)
approximation.  Assuming a very simple model of \bes\ as a bound state
of nonrelativistic \het\ and \hef\, and D as a bound state of $n$ and
$p$, and neglecting nuclear spin forces, we arrive at the following
estimate for the relation between the R1 and R2 cross sections and
those of the $\bes(\gamma,\alpha)\het$ and ${\rm D}(\gamma,p)n$
processes:
\begin{equation}
\label{relation}
\frac{\sigma_{{\rm abs}, i} v }{\sigma_{{\rm photo}, i} c } \simeq \frac{C_i}{4\pi\alpha}\times\frac{m_a^2}{f_d^2},
\end{equation}
where $i=\bes,{\rm D}$ and the coefficients
$C_{\bes}= \frac{64}{3},\, C_{\rm D} =\frac{59}{9}$ reflect spin
combinatorial factors.  The photo-absorption cross section by D is
well-known, while for \bes\ we use recent evaluations
\cite{Cyburt:2008up}. We conclude that $f_d \sim \TeV$ yields both
lifetimes and absorption cross sections in the desired ballpark.

The remaining undetermined parameter is the abundance $n_a$ prior to
decay. It is easy to see that obtaining the correct abundance range
would require some depletion of $a$: despite its small width, $a$ will
get thermally populated during the QCD epoch. We have examined several
ways of depleting its abundance, all of which require additional
particles in the light sector. Disregarding the issue of technical
naturalness of small scalar masses, one can imagine that a coupling to
a nearly massless scalar $s$, $\frac{\lambda}{4}a^2s^2$, mediates the
depletion of $a$ at $T\sim m_a$ via $aa\to ss$. Given the annihilation
cross section $\sigma_{\rm ann} v = \lambda^2/(64\pi m_a^2)$, the
entire range of abundances is covered for
$10^{-5} \lsim \lambda \lsim 10^{-1}$. Alternatively, one can achieve
a similar depletion of $a$ via co-annihilation with another light
species, or via the $3a\to 2a$ process as, {\em e.g.}, in
\cite{Hochberg:2014dra}. More details on viable cosmological models of
ALPs will be provided in~\cite{GPPinprep}.

Scenario B, with unstable particles decaying to massless (or nearly
massless) ALPs, $X_p \to aa$, is even easier to implement.  Consider a
nearly massless ALP $a$, and its progenitor $X_p$ coupled to the SM
via
\begin{equation}
{\cal L}_{XX_p} = A X_p(H^\dagger H)  + B X_p a^2 + {\cal L}_{aq},
\end{equation}
where $H$ is the SM Higgs field. The required abundance of a parent
scalar $X_p$ can be achieved via the ``freeze-in'' mechanism (see,
{\em e.g.}, \cite{Pospelov:2010cw}) by dialing the mixing with the SM
Higgs, $A \sim (10^{-9}-10^{-5})$~\GeV.  The decay of $X_p$ to ALPs is
controlled by the $B$ parameter, and $\tau_{X_p}\sim 10^3$~s is
achieved with $B \sim 10^{-11}$~\MeV.  The nuclear breakup cross
sections due to a massless axion can again be related to the
photo-nuclear cross section \cite{Pospelov:2008jk}.  Performing
calculations similar to (\ref{relation}), we find
\begin{eqnarray}
\frac{\sigma_{{\rm abs}, i} }{\sigma_{{\rm photo}, i}  } \simeq \frac{D_i}{4\pi\alpha}\times\frac{E_a^2}{f_d^2},
\end{eqnarray} 
with $D_{\bes}= \frac{128}{9},~D_{\rm D} =\frac{118}{27}$. In calculating the impact on BBN in this scenario, 
we account for the redshifting of $E_a$ from $m_{X_p}/2$ to R1 and R2 thresholds. 

\begin{figure}
\centering
\includegraphics[width=\columnwidth]{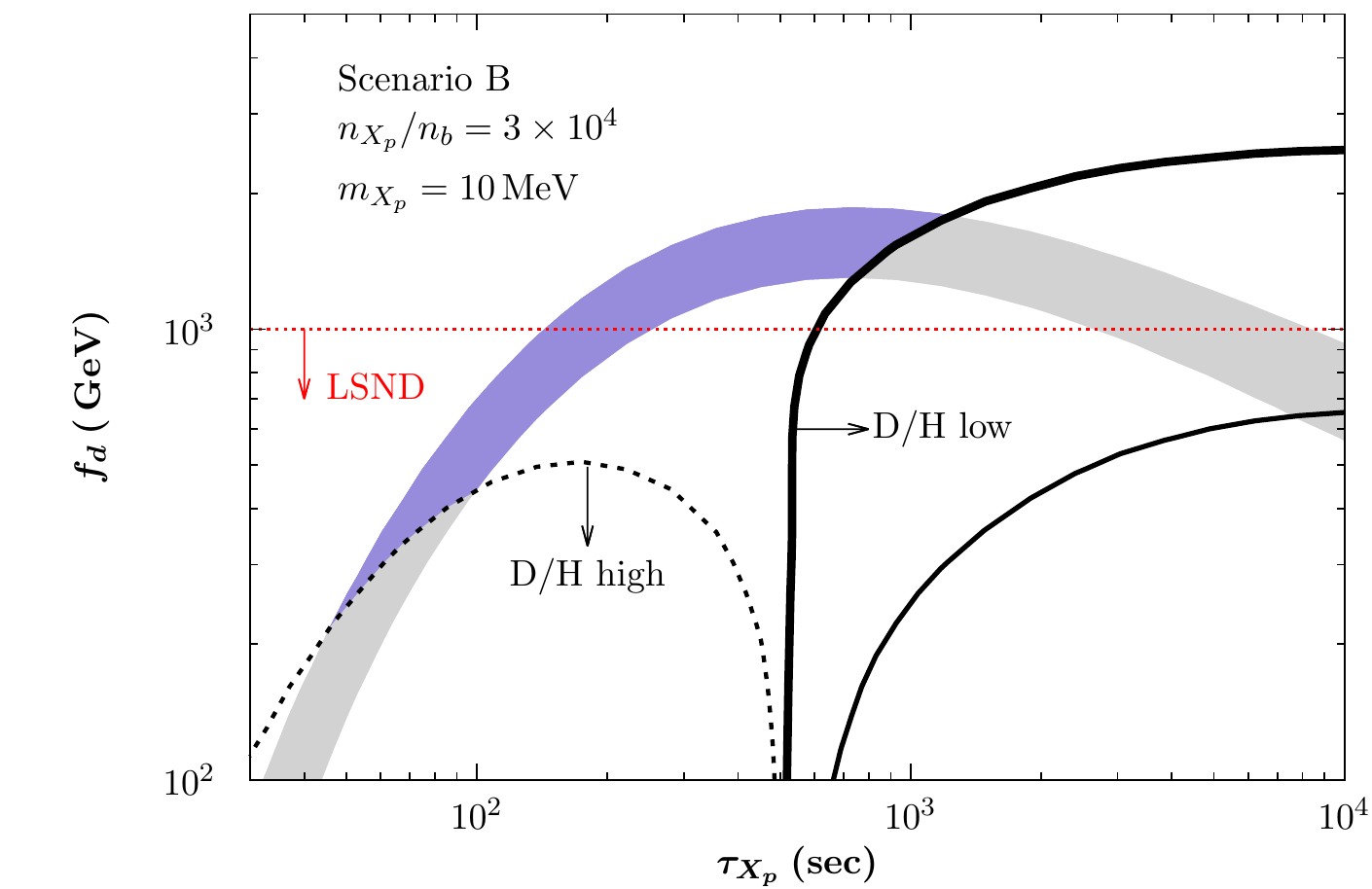}
\caption{\small Lithium solution by ALPs that are injected from a
  progenitor state $X_p$ with mass $m_{X_p} = 10\,\MeV$. The LSND
  sensitivity-line is fixed, but all other contours can move
  vertically by adjusting the $X_p$ initial abundance
  $n_{X_p}/n_b $.}
\label{fig:planb}
\end{figure}

\paragraph{Searching for hadronic ALPs at the intensity frontier.}
Our proposal for the lithium reduction mechanism involves light
particles in the several MeV range, but with rather small
couplings. Such particles are being searched for at intensity frontier
experiments \cite{Essig:2013lka}. To better define the parameter space
of interest, we take Scenario~B, and vary $\tau_{X_p}$, and $f_d$, by
fixing a fiducial value of $n_{X_p}$, the $X_p$ abundance prior to
decay. The results are shown in Fig.~\ref{fig:planb}. The ``pile-up''
from redshifted $X$ results in sensitivity to lifetimes before the end
of the D-bottleneck, $\tau_{X_p} < 100\,\sec $; with
$n_{X_p} \sim 10^4 n_b$ a depletion of lithium by a factor of a few
is possible with $f_d \sim \TeV$.

Next, we estimate the expected signal in beam dump experiments such as
LSND \cite{Aguilar:2001ty}. The ALP-production in $p$-nucleus
collisions is followed by the scattering/absorption of $a$ by nuclei
of the target. We assume that the number of produced ALPs scales with
the number of produced $\pi$-mesons as
$N_a \sim (f_\pi/f_d)^2\times N_\pi$. Concentrating on the photon
production in the $p(a,\gamma)p$ process, we estimate its cross
section~\cite{Pospelov:2008jk} as
$\sigma_{ap} \sim \alpha (E_a/f_d)^2m_p^{-2}\sim
(100~\MeV/f_a)^2\times 10^{-29} {\rm cm^2}$,
where $E_a \sim 200~\MeV$ is a typical energy of produced mesons and
ALPs \cite{Auerbach:2003fz}. The estimated number of events
\begin{equation}
  N_{\rm events} \sim \frac{N_aN_{ p} \sigma_{a p}}{4\pi L^2} \sim 6 \times  \left( \frac{\rm TeV}{f_d} \right)^4
\label{LSND}
\end{equation}
should be compared to the number of prompt energetic events in the
detector, $O(10)$, which implies a sensitivity up to $f_d \sim 1~\TeV$.
Here, $L= 30~{\rm m}$, $N_\pi \sim 10^{23}$ and
$N_p = 6.7\times 10^{30}$ is the number of target protons inside the
fiducial volume. One can see, Fig.~\ref{fig:planb}, that---depending
on the assumed abundance of the progenitor $X_p$---LSND can probe
large fractions of relevant parameter space; further significant
improvements can be achieved by deploying beam dump experiments next
to large underground neutrino detectors~\cite{Izaguirre:2015pva}.

\paragraph{Conclusions.}  We have shown that particle physics
solutions of the cosmological lithium problem are far from being
exhausted. Light, very weakly interacting particles with energy or
mass of $\sim 10~\MeV$ and lifetimes of $O(10^3)$ seconds can deplete
\bes+\lisv\ without affecting other elements.  This is because, unlike
in many weak-scale solutions, the suggested mechanism does not inject
any new neutrons into the primordial medium, and operates either via
direct destruction of \bes, or through its indirect reduction via
neutrons that are temporarily ``borrowed'' from deuterium. A variety
of particle physics realizations of this idea is possible, and in
particular ALPs with small couplings to $d$-quarks represent a clear
target of opportunity for upcoming searches at the intensity frontier.

\paragraph{Acknowledgements.}AG and JP are supported by the New
Frontiers program of the Austrian Academy of Sciences. The work of MP
is supported in part by NSERC, Canada, and research at the Perimeter
Institute is supported in part by the Government of Canada through
NSERC and by the Province of Ontario through MEDT.}

\bibliography{biblio}

\end{document}